# High-Resolution Probing of Molecular Junctions: Vibrational Fingerprinting and Parameter Extraction via Current Noise Spectroscopy

**Rani Arielly**[a]

[a] Institute of Agricultural and Biosystems Engineering, Agricultural Research Organization - Volcani Institute, Rishon LeZion, 7505101, Israel. Email: rania@volcani.agri.gov.il

Corresponding author: Rani Arielly (rania@volcani.agri.gov.il)

## Abstract

The precise realization of molecular electronic devices requires a comprehensive understanding of charge transport mechanisms and the specific interplay between electronic and nuclear degrees of freedom. While average current measurements ($I-V$ characteristics) and conventional Inelastic Electron Tunneling Spectroscopy (IETS) offer valuable insights, they are fundamentally limited by temperature-dependent line-width broadening. This study presents a high-resolution spectroscopic methodology utilizing suspended-wire molecular junctions (SWMJs) based on self-assembled monolayers (SAMs) of 1-decanethiol (C10) and 1,1',4',1"-terphenyl-4-thiol (TPT). By systematically probing the voltage-dependent current noise ($\Delta I$), we demonstrate that electronic noise spectroscopy circumvents thermal degradation by probing transition rates between vibrational manifolds rather than simple additions of conductance channels, which enables sub-thermal feature mapping. Leveraging a fast-convolution-based Landauer-Büttiker transport model fitted to experimental data, we map complex vibrational manifolds, including high-energy overtones. This allows for the direct extraction of crucial nanoscale molecular parameters, including mode energies, anharmonicities ($x_e$), dissociation energies ($D_e$), and local environment reorganization energies ($E_r$). These parameter-dense noise signatures act as a unique molecular fingerprint, establishing noise spectroscopy as a highly sensitive platform for chemical sensing and discrimination in advanced quantum devices.

## 1. Introduction

The realization of functional molecular electronic devices requires a precise understanding of charge transport mechanisms and the specific interplay between electronic and nuclear degrees of freedom at the nanoscale (Evers et al., 2020; Galperin et al., 2008; Xu et al., 2024). The organic molecule-metal interface is incorporated in many devices, yet it remains poorly understood in some aspects, particularly the effect of the conjugated system on the vibrational spectrum (Galperin et al., 2004; Troisi and Ratner, 2005). While average current measurements ($I-V$ characteristics) often obscure the rich dynamics inherent to transport, advanced techniques such as Inelastic Electron Tunneling Spectroscopy (IETS) have become important for identifying molecular vibrational signatures. The advantage of IETS lies in its non-selection-rule nature as opposed to infra-red (IR) or Raman spectroscopies. However, standard IETS, which relies on measuring $d^2I/dV^2$ directly, is frequently limited by the fact that the vibration peak's line width is dependent on the temperature. This necessitates ultra-low temperatures to resolve individual modes, thus limiting IETSs utility as a general-purpose spectroscopic tool (Hansma, 1977; Lambe and Jaklevic, 1968; Reed, 2008; Troisi and Ratner, 2006).

Electronic noise spectroscopy can circumvent these limitations by looking at the intrinsic dynamics of the system, regardless of the curvature of the Fermi-Dirac distribution in the leads. Current noise, or the correlation of current fluctuations, provides access to information unavailable through conductance alone (Yuan et al., 2021). While shot noise measurements have proven to be a powerful method for evaluating Fano factors and the number and transparency of channels for charge transmission (Djukic and Van Ruitenbeek, 2006; Galperin et al., 2008), recent theoretical and experimental work suggests that inelastic scattering processes - specifically the interaction between tunneling electrons and molecular vibrations - imprint distinct signatures on the current noise spectrum (Haupt et al., 2010; Tsutsui et al., 2010). Instead of probing the increase in conductance channels like IETS, noise spectroscopy probes the noise power, which is dominated by the transition rate between vibrational manifolds. This effectively creates sub-thermal probing.

In this work, we present a high-resolution spectroscopic study of suspended wire molecular junctions based on self-assembled monolayers (SAMs). By systematically analyzing the voltage-dependent current noise, we demonstrate that noise spectroscopy yields peak-shaped spectral data in much finer detail than conventional IETS measurements. Notably, the observed spectral resolution exceeds the theoretical limits of standard inelastic tunneling models at the measured temperature (Lambe and Jaklevic, 1968), potentially extending the operational regime of molecular spectroscopy toward higher temperatures. This ability to resolve such fine spectral features suggests a direct application as a molecular sensor.

We devise a theoretical model and by fitting the model to our experimental data, we are able to report the resolution of complex vibrational manifolds, allowing for the precise deduction of mode energies, anharmonicities, dissociation energies, and reorganization energies for different modes of the molecular species. These parameters provide a molecular signature and add to the uniqueness of the device as a molecular sensor.

Because the noise spectrum is acutely sensitive to the local environment and vibrational states of the molecule, this technique offers a promising platform for detecting minute chemical changes. By using these noise signatures as a molecular fingerprint, this method could allow for a molecular sensor capable of discriminating between molecules that appear identical to standard conductance probes.

## 2. Experimental Methods & Observations

“Suspended-wire” molecular junctions (SWMJ) were fabricated by trapping Au nanowires, ~200 nm in diameter, capped with self-assembled monolayers (SAM) of either 1-decanethiol (C10), or 1,1′,4′,1″-Terphenyl-4-thiol (TPT), onto lithography-defined Au leads using a dielectrophoresis technique (see Figure 1) (Noy et al., 2010). The nanowires which are completely covered with a molecular layer could potentially form two molecular junctions in each SWMJ, only one junction per suspended nanowire (and a short on the other end), and a complete short circuit.

IETS measurements were done under vacuum at the boiling temperature of liquid nitrogen at ambient pressure ($T_{LN}$) using a standard lock-in technique, with amplitude of 15 meV at frequency of ~1 kHz, revealing typical vibrations at both bias polarities (Okabayashi et al., 2008). These measurements were helpful in sorting out the different types of devices, leaving only the single junction type device for further analysis (see Figure 1).

Electrical noise measurements were conducted under the same conditions as the IETS measurements, by supplying a series of bias voltages ($V$) to the metallic leads and acquiring ~300 current measurements at each bias step at a sampling frequency of 17 KHz. Due to the noise in bias, measurements were sorted according to the monitored bias values, which became sharp step like behavior due to the measurement resolution. From each step, a mean value ($I$) and standard deviation ($\Delta I$) were calculated. Both were smoothed (nearest neighbor, 50 points = 15.9 mV window).

Typical examples of current measurements for both types of devices are shown in Figure 2. Except for slight asymmetries in the curves due to the asymmetry in the molecular structure of our SWMJs, it seems like they don't provide us with much information. This, however, changes when examining $\Delta I$ as a function of $V$. Typical examples for $\Delta I$ are presented in Figure 3, each corresponding to a $I(V)$ curve in Figure 2 in the order of appearance. As opposed to the $I(V)$ curves, the $\Delta I(V)$ curves feature non-monotonous, numerous peak features. Since our general notion was that multiple features in this energy range (<0.5 eV) should originate from vibrational related phenomena, and these should be symmetric in bias polarity (at least in an ideal system), the peak features were tested for peak/onset bias value and bias symmetry. It was also clear that the highest biased peaks won't fit any known vibrational energy since these are maxed at $\sim 3000 - 3500\ cm^{-1}$ for O-H stretching, so the possibility of vibrational overtone was considered. Such a test is visible at the bottom of each panel in Figure 3 which include some initial (not fitted) mode Assignments that are based on the known modes for both adsorbed molecules and $H_2O$ with additional peaks assigned with overtones of the base modes, calculated with first order anharmonicity (Frisch et al., 2016; Gordon and Schmidt, 2005; Pretsch et al., 2009; Schmidt et al., 1993; Wallace, 2018). For the C10 junctions, we were able to assign all peaks using only the base frequencies (blue markers) and the first overtones (green markers), while for the TPT junctions the second overtones were also needed (red markers). Since analyzing over a hundred of these curves individually seemed cumbersome, and since we are only interested in typical values for each type of device, typical $I(V)$ and $\Delta I(V)$ curves were calculated by harmonically averaging (Ferger, 1931) the individual measured curves in bins of a uniform $V$ space (1 mV resolution). The final $I(V)$ and $\Delta I(V)$ curves are shown in Figure 4.

Immediately it was clear that $\Delta I(V)$ contain non-uniform peak features in spite of the large number of averaged measurements (96 and 54 for the C10 and TPT junctions respectively). The peak features' energies were found to correspond with known vibrational modes, some appearing in the IETS measurements and some not clearly resolved. Although others have measured and modeled vibrations related features in electrical systems (Haupt et al., 2010; Tsutsui et al., 2010), the appearance of peak features was surprising and lacking any existing explanation and therefore a new model was formalized.

## 3. Theoretical Framework and analysis

### *3.1. Current passing through a multimode molecular device*

The current is described by the Landauer-Büttiker formula with added dependencies on time ($t$) and voltage in the transmission term due to the dynamical nature of the molecules inside the monolayer

$$I(t) = 2Ne/h \int dE\big(f_L(E,V,T) - f_R(E,V,T)\big)\tau(E,V,t) \tag{1}$$

where $e$ is the electron's charge, $h$ is the Planck constant, $N$ is the number of connecting molecules, $E$ is the energy, $T$ is the temperature, $f_{L/R}$ are the Fermi-Dirac distributions of the left (L) and right (R) electrodes and $\tau$ is the transmission function. We can claim that the current measured during the sampling time is composed out of many contributions: a vibrational mode may be excited or not, and in each vibrational state the molecule's electron affinity also varies in time differently when the molecule moves along the vibration coordinate, $x$ (Yoshida et al., 2021). Considering a large number of passing electrons, we can take these into consideration by writing $\tau(E,V,t)$ as a series of contributions, weighted by the molecule's probabilities for being excited and at specific positions. For a single mode, $n$, this will give

$$I = 2Ne/h \int dE\big(f_L(E,V,T) - f_R(E,V,T)\big)\left\{P_n^g(V)\sum_{x_n\in\mathbb{R}_n^g} P_{osc,n}^g(x_n)\tau(E,x_n) + P_n^{ex}(V)\sum_{x_n\in\mathbb{R}_n^{ex}} P_{osc,n}^{ex}(x_n)\tau(E,x_n)\right\} \tag{2}$$

where $P_n^{g/ex}(V)$ are the probabilities for the molecules to be in the vibrational ground/excited state of the $n$ mode, $\mathbb{R}_n^{g/ex}$ are the coordinate spaces for the molecular movement in the ground and excited states, $P_{osc,n}^{g/ex}(x_n)$ is the probability for the molecule being at specific location, $x_n$, along the vibration coordinate and $\tau(E,x_n)$ is the transmission function of the molecule in a specific configuration. For a molecular system which can have multiple vibrational modes, which can be excited or not, we can generalize this and get

$$I = 2Ne/h \int dE\big(f_L(E,V,T) - f_R(E,V,T)\big)\sum_i P_i(V)\tau_i(E) \tag{3}$$

where $P_i(V)$ are the probabilities for the molecule to be in specific configuration, $i$, for all the modes, and $\tau_i(E)$ is the effective transmission function for that configuration.

### 3.2. State probabilities and transition rates

Considering low population density for the excited states, the probabilities for mode excitations are calculated by considering a rate equation for the number of molecules that are in an excited state, $N_{ex}$, and those that are in the ground state, $N_g$ (omitting here the mode designation, $n$)

$$dN_{ex}/dt = -dN_g/dt = k_{ex}N_g - k_g N_{ex} \tag{4}$$

which is determined by the rates of excitations, $k_{ex}$, and decays, $k_g$. The probability for having an excited molecule is the ratio between the number of excited molecules and the total number of molecules ( $P^{ex}(V) = N_{ex}/N$ ). By solving the rate equations in equilibrium (=0) we get the number of molecules in the vibrational ground and excited states, and thus, the probabilities for having molecules in the excited and ground states.

$$\begin{gathered} P^{ex} = k_{ex}/\left(k_{ex} + k_g\right) \\ P^{g} = k_g/\left(k_{ex} + k_g\right) \end{gathered} \tag{5}$$

In order to calculate the rates for excitation and decay, we need to consider the typical time for the mode's decay, $t^{decay}$, which is inversely proportional to the decay rate. In order to calculate the excitation rate, we will again use the Landauer-Büttiker formula and modify it to only "count" tunneling electrons with sufficient energy in order to excite the molecular vibrational

$$\begin{aligned} k_{ex} = 2\xi/h \int dE \{ & \Theta(eV - E_{vib})\left(f_L(E,V,T) - f_R(E - E_{vib},V,T)\right) \\ & + \Theta(-eV - E_{vib})\left(f_R(E,V,T) - f_L(E - E_{vib},V,T)\right)\}\tau(E) \end{aligned} \tag{6}$$

where $\Theta$ is the Heaviside function, $E_{vib}$ is the energy of the vibration and $\xi$ is the probability for a vibrational excitation to occur as a result from energy transfer from a tunneling electron. There is also the thermal excitation rate, however this will not contribute to voltage dependent features (also the thermal energy in our experiment is much lower than the lowest measured vibrational mode) and may be discarded.

Considering a molecular state, $i$, that includes the ground/excited states for all $M$ possible modes at some possible combination, the probabilities $P_i(V)$ are calculated as the product of the mode state probabilities. Given a specific energy mode, temperature, and bias potential, $P_i$ depends on two adjustable parameters - $\xi$ and $t^{decay}$. It is important to mention, however, that it is not possible to have a unique set of solutions with this model for each of the parameters. If we express $k_{ex}$ as $k_{ex} = \xi J$, then the probabilities are

$$\begin{gathered} P^{ex} = (1 + (J\xi t^{decay})^{-1})^{-1} \\ P^{g} = (J\xi t^{decay} + 1)^{-1} \end{gathered} \tag{7}$$

thus, the only real parameter here is $s \equiv \xi t^{decay}$.

### 3.3. Effect of the molecular motion on the transmission function

In order to evaluate $\tau_i(E)$, we consider the molecular motion as of a quantum harmonic oscillator. In our case, we are interested in the ground, excited, 1$^{st}$ overtone and 2$^{nd}$ overtone states (as estimated in Figure 3), and we can calculate configuration probabilities terms in Eq. (2) for these state's using their known wavefunctions ( $\sum_{x_n \in \mathbb{R}_n^l} P_{osc,n}^l(x_n)\tau(E,x_n) = \sum_{x_n \in \mathbb{R}_n^l} |\psi_l(x_n)|^2 \tau(E,x_n)$ ). Following Eq. (3), consider molecular state $i$, which represents a specific combination of excitation states across all $M$ modes. If molecular movement skews the orbitals, the transmission function $\tau_i(E)$ should be evaluated by redistributing probabilistic values between modes. This is achieved by mapping the non-disturbed transmission function across the molecular coordinate and probability space (Ruhoff, 1994).

$$\tau_n(E) = \int_{-\infty}^{\infty} dx_n P_{osc,n}^l(x_n) \tau_{n-1}(E'(x_n)) \tag{8}$$

$\tau_i$ is achieved once all modes have been accounted for. i.e. $\tau_i \equiv \tau_M$. $\tau_0(E)$ is the transmission function of the molecule without the effects of molecular vibrations. For simplicity, it can be described by the following Lorentzian function

$$\tau_0(E) = (\gamma/2)^2 \left[ \left(E - E_g\right)^2 + (\gamma/2)^2 \right]^{-1} \tag{9}$$

where $E_g$ is the energy difference between the electrode's fermi level and the nearest molecular orbital and $\gamma$ is the energy width of that orbital. $E'(x)$ is the modified energy due to the atomic movements across the coordinate $x$ (Nocera et al., 2011)

$$E'(x) = E - Am\omega^2 x^2/2 \tag{10}$$

where $m$ is the molecular oscillator mass, $\omega$ is its base frequency, and $A$ is the electrostatic reduction factor that accounts for mirror charges that would partially prevent the electron cloud from distorting (Perrin et al., 2013; Xie et al., 2021). In a multilayer device, $A$ can be set individually for each phase based on its molecular neighborhood. It is also possible to write this integration in energy terms ($U = E - E'(x)$). $\tau_n(E)$ then becomes

$$\tau_n(E) = \int_{0}^{\infty} dU P'^l_{osc,n}(U) \tau_{n-1}(E - U) = (P'^l_{osc,n} * \tau_{n-1})(E) \tag{11}$$

This form for $\tau_n(E)$ is much preferable since fast convolution implementations are available which shortens calculation time by a factor of ~2000. Additional advantage is the elimination of the mass term which simplifies the problem. However, it is still difficult to calculate $P'^l_{osc,n}(U)$ when $U \ll 1$. Following the convolution theorem, if we mark the Fourier transform of function $f$ as $\mathcal{F}\{f\}$, then

$$\tau_n(E) = \mathcal{F}^{-1}\left\{\mathcal{F}\left\{P'^l_{osc,n}\right\}\mathcal{F}\{\tau_{n-1}\}\right\} \tag{12}$$

The original normalization of $\tau_0$ will be conserved.

It is also possible to calculate the final transmission function, $\tau_i$, with a single step

$$\tau_i(E) = \tau_M(E) = \mathcal{F}^{-1}\left\{\mathcal{F}\left\{P'_{i,M}\right\}\mathcal{F}\left\{P'_{i,M-1}\right\} \cdots \mathcal{F}\left\{P'_{i,1}\right\}\mathcal{F}\{\tau_0\}\right\} \tag{13}$$

where $\mathcal{F}\left\{P'^l_{osc,n}\right\}$ and $\mathcal{F}\{\tau_0\}$ can be analytically calculated

$$\mathcal{F}\{P'^{0}_{osc,n}\} = B_n^{-1/2}$$
$$\mathcal{F}\{P'^{1}_{osc,n}\} = B_n^{-3/2}$$
$$\mathcal{F}\{P'^{2}_{osc,n}\} = B_n^{-1/2}[3B_n^{-2} - 2B_n^{-1} + 1]/2 \tag{14}$$
$$\mathcal{F}\{P'^{3}_{osc,n}\} = B_n^{-3/2}(5/2\, B_n^{-2} - 3B_n^{-1} + 3/2)$$
$$\mathcal{F}\{\tau_0\} = \gamma\pi \exp\left(-2\pi i y E_g - \gamma\pi|y|\right)/2$$

where $y$ is the energy conjugated Fourier space parameter and $B_n = 1 + A\pi i y\hbar\omega_{\mathrm{n}}$. Thus, we have eliminated the divergence problem in $P'^{l}_{osc,n}(U)$, and simplified the calculation even more by using a fast Fourier transform a single time, instead of convolving for each mode.

the term for $k_{ex}$ (Eq. (6)) is dependent on $\tau(E)$ which is expected to change for each newly available mode. The overall effect is to push $\tau(E)$ to higher energies and decrease the current, which itself is the propagator of this model, thus, $k_{ex}$ and $\tau(E)$ should converge. To reach the real values, one has to recalculate $k_{ex}$ and $\tau(E)$ until convergence is achieved. However, since the maximal changes in current were not significant, this recalculation was not done, as it was cumbersome to implement and gave no substantial insight.

### *3.4. Noise sources and peak features*

When examining the above expressions, several parameters can be identified as error (noise) sources for the electrical current: (1) The uncertainty in $t^{decay}$ is mainly set by the low frequency movement of the monolayer that is coupled to the suspended gold wire which is known to oscillate at ~1 MHz (Noy et al., 2010; Vijayaraghavan et al., 2007) – the far edge of the expected $t^{decay}$ scale which is in the nanoseconds(Spears, 1971) so this should manifest as ~1% change. (2) $\xi$ should be set by the scattering cross section which is dependent mainly on features of the monolayer like angles and spacings which should remain constant in a monolayer, but can change due to surface impurities which our measurement is sensitive to, and considering our measurement's quality we can cap these changes at ~1%. (3) The number of connecting molecules in the monolayer ($N$) may fluctuate. (4) Fluctuating temperature may be a source for electrical noise. (5) The voltage bias may also fluctuate around a set value. The bias source used has fluctuations that were measured to be 0.1-0.4 mV (standard deviation). Following Eq. (3), the combined error term is

$$\Delta I = \sqrt{\Delta {I_N}^2 + \Delta {I_s}^2 + \Delta {I_T}^2 + \Delta {I_V}^2} \tag{15}$$

where the different error terms are derived by differentiation to the error parameters and taking the root sum of squares across different modes, as if dealing with uncorrelated systems, or a series of events without memory. In calculating $df_{L/R}/dV$ we assumed symmetrical potential change between the electrodes. Since both $f_{L/R}$ and $P_i(V)$ are bias voltage dependent, the expression for $\Delta I_V$ contain two elements. The second part of $\Delta I_V$ ($\Delta I_{V2}$), which includes the derivative of $P_i(V)$, gives peak features while the first part ($\Delta I_{V1}$) gives a monotonic slope. These four electrical current error terms are plotted in Figure 5 for a single mode using $T = 80\ K$, $\Delta T = 1\ K$, $N = 100$, $\Delta N =$

1, $s = 1000\,psec$, $\Delta s = 10\,psec$, $\Delta V = 0.3\,mV$, $E_g = 2\,eV$, $\gamma = 0.1\,eV$, $A = 0.2$ and $\omega = 3e14\,rad/sec$ ($1600\,cm^{-1}$). These values were chosen in order to show the voltage dependence of the four error terms as clear as possible in a single graph while being somewhat realistic. $\Delta I_T$ values were increased by a factor of 200 for the same reason. It is clear, when searching for peak features in the electrical noise, that the dominant term that contribute such a feature is $\Delta I_V$, and the next dominant one is $\Delta I_N$ which adds a relatively linear slope. These two behaviors fully describe our measurements' features, and thus the error term is approximated as

$$\Delta I \cong \sqrt{{\Delta I_N}^2 + {\Delta I_V}^2} \tag{16}$$

which is shown as a dashed line in Figure 5 after being smoothed similarly to our measurements. Due to the inability to account for the lowest energy modes in our analysis, the acknowledgement of other, featureless, error terms in our derivation, and some inherent electrical noise in our instrumentation, we also add a constant noise term, $\Delta I_0$, to Eq. (16).

For a device containing multiple successive molecular layers, the current term will include the product of all the probabilities for these layers ( $\prod_l \sum_i P_{i,l}(V)\,\tau_{i,l}(E)$ ), thus $\Delta I_{V2}$ will include contributions from each of these layers combined in the same uncorrelated way a before. In our junction's fabrication process, electrochemically grown gold wires are covered in SAM and placed on a gold surface. Since we consider our fabrication less than perfect, we expect water residues on our non-protected gold surfaces, and thus, a "gold electrode-SAM-water-gold electrode" layered structure.

When changing the parameter $s$ due to a change in the damping time of the mode or a change in the probability for exciting the mode, the peak feature in $\Delta I_V$ becomes sharper with increasing $s$ and broadens into a step like feature with decreasing $s$. This is shown in Figure 6. When dealing with multimode systems, the only parameters specific to a mode are $\omega$ and $s$, however, since $\omega$ is set by the location of the peak, when fitting this model to a noise measurement, the only parameter available to determine a peak's amplitude is $s$, which adds to the uniqueness of the solution.

## 4. Results and Discussion

### *4.1. Model fitting and parameters*

Figure 4 shows fits of the model to $I(V)$ and $\Delta I_V(V)$ for 2 types of devices (C10 and TPT). The set of parameters that produces the best fit is shown in Table I. $I$ and $\Delta I_V$ were both smoothed like the experimental data with a constant smoothing window (15.9 mV) but with an added gradient that increased the smoothing window at higher bias values. This gradient is attributed to non-uniformities in the electrostatic potential drop across the device when applying the bias voltage. This is common in these type of devices and becomes more severe when the bias voltage rises (Nitzan et al., 2002). The first thing that is apparent from these results is the richness of the measured graphs compared with other methods (IETS, IR, Raman) without losing quality compared to IETS and even surpassing it, considering the simple measurement method and the not so low liquid nitrogen temperature. The second is the assortment of extracted parameters, many of not known until now in an organic molecule / metal system. While the measurement method was not novel (Tsutsui et al., 2010), it's application in our SWMJ devices opened a window for new physics since the previously available models (Galperin et al., 2008; Haupt et al., 2010) failed to describe our measurement's peak features, due to the assumption that all of the important physics is due to the addition of conductance channels.

When examining the results in Figure 4 and Table I one may wonder how were the vibration modes assigned to the graphs features and if the assignments here are the only possibility. On top of that, although the mode's energies are as expected from literature in respect to one another (Pretsch et al., 2009; Wallace, 2018), their absolute energies are shifted as a whole. The logic behind the assignments here is: first – we know that the features we measure are vibration related due to previous evidence (Tsutsui et al., 2010), second – the highest energy peaks are far too energetic to be associated with any known basic vibration, and therefore it should be from overtones, third – the overtones shouldn't necessarily be integer multiples of the base frequency and some anharmonicity may be in effect, fourth – possible fabrication contaminations should be taken into account (in our case water and gold cyanide were considered but only the first was needed for a good fit). The fact that we used the averaged spectrum also made the use of $2^{nd}$ overtones redundant. These guidelines were combined with known infra-red data (Pretsch et al., 2009; Wallace, 2018), and molecular quantum chemistry programs (Frisch et al., 2016; Gordon and Schmidt, 2005; Schmidt et al., 1993) in order to assign the different peaks in our measurements. Some discrepancy exists between the reference sources, and we believe that this is due to the different anharmonicity constants that exist in different systems/environments. This is discussed hereafter.

The fitted values for $E_g$ and $\gamma$ correspond to the known values for the SAM molecules and water, especially when attached to metallic surfaces (Arielly et al., 2014; Guo et al., 2014; Seminario and Yan, 2005; Wu et al., 2013). The effect of the latter becomes more apparent in the $A$ parameter of the SAM layer Vs. the water layer which has lower $A$ values. As stated above $A$ is the electrostatic reduction factor that accounts for mirror charges that would partially prevent the electron cloud from distorting. Since this should depend on the neighborhood of the molecule, we expect that when a small molecule like water is adsorbed on a non-smooth metallic surface, this would result in very low $A$ values compared to a molecule that is ~1 nm partially distant from the metallic surface. The effect of mirror charges was also studied in the field of plasmonics in small gaps experimentally and theoretically (Arielly et al., 2011; Kurokawa and Miyazaki, 2007; Miyazaki and Kurokawa, 2006). On that note we can see that the TPT molecule is more conjugated with the gold compared to the C10 molecule, evident by a lower $A$ value which we may interpret as shorter gap, and a higher $N$ value which is a wider base for the junction structure. $\Delta N/N$ values are lower for higher $N$ values, which may mean stable molecular connections deep inside the monolayer and random disconnects at its edges.

### *4.2. Parameters of the anharmonic potential*

For a specific mode, the energy levels of the harmonic oscillator that was described before, considering only first order anharmonicity with parameter $x_e$, are

$$G(l) \cong \tilde{\nu}_e(l+1/2) - \tilde{\nu}_e x_e (l+1/2)^2 \quad (17)$$

where $\tilde{\nu}_e = \hbar\omega$. Our measurements were fitted to the transitions between the ground states and the first and second excited levels. The energies of these transitions are

$$\begin{aligned} \Delta G_1 &= G(1) - G(0) = \tilde{\nu}_e - 2\tilde{\nu}_e x_e \\ \Delta G_2 &= G(2) - G(0) = 2\tilde{\nu}_e - 6\tilde{\nu}_e x_e \end{aligned} \quad (18)$$

Comparing these expressions to the measured peak locations in Table I can allow us to calculate $\tilde{\nu}_e$ and $x_e$

$$\begin{aligned} \tilde{\nu}_e &= 3\Delta G_1 - \Delta G_2 \\ x_e &= (2\Delta G_1 - \Delta G_2)/2\tilde{\nu}_e \end{aligned} \quad (19)$$

Using these values, it is possible to use the Birge-Sponer extrapolation method. The vibrational energy differences are

$$\Delta\nu = \tilde{\nu}_e - 2\tilde{\nu}_e x_e (l+1) \quad (20)$$

Which should converge to zero around

$$l_c + 1 = (2x_e)^{-1} \quad (21)$$

Thus, calculating the area under the curve from $l = 0$ up to $l_c$ should give us the dissociation energy

$$D_0 = (l_c+1)\tilde{\nu}_e/2 = \tilde{\nu}_e/(2x_e) \quad (22)$$

and the depth of the electronic potential

$$D_e = D_0 + \tilde{\nu}_e/2 = \tilde{\nu}_e(x_e^{-1}+1)/2 \quad (23)$$

The values of $\tilde{\nu}_e$, $x_e$ and $D_e$ for the different modes were calculated Based on Eq. (19),(23) and Table I and are presented in Table II (besides two low energy modes that were left out due to high uncertainties). When examining these results, one can recall the well-known gaseous iodine results for comparison (McNaught, 1980), where $x_e \cong 0.003, 0.008$ and $D_e \cong 12000, 4000\ cm^{-1}$ for the ground and excited electronic levels respectively. While our values of $D_e$ are quite similar, with an average of 10654(73) $cm^{-1}$, the $x_e$ values are quite different, with an average of 0.1871(13). Straightforwardly speaking, we can sum this by stating that no matter how energetic is the vibration mode, the dissociation energy would remain at the same order of magnitude, and only the anharmonicity will change to compensate. Another conclusion that we can draw from this is on the low stability of these devices when exciting them optically. Occasionally one would design his experiment by considering a device based on a molecule with known HOMO-LUMO gap, and irradiating it with a light source with the same photon energy while measuring the electrical response. The average value of $D_e$ we calculated translates to 1.51 eV or 821 nm, which mean that unless the light source is deep inside the infrared spectral range, the photon energies are more than enough to cause atomic dissociation of the molecule and break the device. As mentioned before, we used molecular quantum chemistry programs in order examine the full set of possible vibration modes for each molecule, but when doing so, more computationally complex aspects like anharmonicity were not given a full treatment, which resulted in a spectrum where all modes were present but with energies much higher than the ones available at the literature (Pretsch et al., 2009; Wallace, 2018). When examining the results for $\tilde{\nu}_e$ in Table II, it becomes clearer since when $x_e \to 0$ then $\Delta G_l \to l\tilde{\nu}_e$.

### *4.3. Mode decay rates and reorganization energies*

Using an acceptable value for $\xi \cong 0.01$ (Scalapino and Marcus, 1967), we calculated the decay times of the different modes, and by using the modes' wavenumbers, also the number of cycles the vibrations went through before decaying. These values vary by some orders of magnitude from one another, but are normally distributed on a log scale very nicely with average values of $\overline{\log_{10}(t_{decay})} = -7.599(18)$ and $\overline{\log_{10}(N_{decay})} = 6.220(19)$ which seem reasonable (Spears, 1971).

It is possible to evaluate the reorganization energies ($E_r$) for excitations in the different modes by considering the probability to cross between the two states that are described by Marcus parabolas. Such a cross can be initiated by the atomic movements themselves, and so it should result in a rate proportional to the wavenumber, i.e., $\nu c$. This rate, however, would diminish due to the finite reorganization energy which contributes to an activation energy of $E_A = E_r/4$ (when there is no significant energy difference between the two states), giving a Boltzmann factor of $\exp(-E_A/(k_B T))$ that results in the measured rate of $t_{decay}^{-1}$. Therefore, by again using $\xi \cong 0.01$, we can calculate the reorganization energies with

$$E_r = 4k_B T \ln\left(t_{decay}\nu c\right) \tag{24}$$

$E_r$ values for the different vibration modes in the two types of devices are presented in Table III. It is clear that for all modes $0.329 \leq E_r \leq 0.558\ eV$ which is a relatively narrow range that is also with good agreement with the value in (Arielly et al., 2014) which was measured in the same type of devices and at the same temperature, but with different monolayer that exhibited redox reactions. That redox reaction has an inner-sphere reorganization energy of $\sim 0.03\ eV$ (Li et al., 2010) and we can assume that this value should be lower when not charging the molecule, so with an average value of $\bar{E}_r = 0.45828(62)\ eV$, we can conclude that most of the reorganization energy is outer-sphere, and due to the lack of solvent in our system, originates from neighboring molecules in the monolayer, or charges within the metallic electrodes.

### *4.4. Device and model uniqueness*

Lastly, we would like to discuss why our case requires a special treatment, since we cited a previous, similar, experimental research (Tsutsui et al., 2010) that seemed to agree with existing models. The IETS model was first proposed with solid state materials at mind (Lambe and Jaklevic, 1968). In these devices vibrations also occur, but we claim that the time dependent change in the transmission function during the period of the vibrations is much smaller than in our devices (we recently demonstrated the time dependence of this change (Arielly et al., 2017)) due to stabilization of the atoms by the lattice. In our devices such stabilization occurs but only partially (as evident by the $A$ parameter) and even between organic molecule devices it is dependent on the specific type of the device that can provide different surroundings for the molecules. Thus, the molecules in our devices are in that sweet spot that holds them in the right amount in order to produce obvious peak features in the current noise, instead of relying on inconsistent numerical calculations to produce them. This, together with the abundance of parameters that can be extracted from the model that provide a molecular signature, offers a promising platform for a molecular sensor.

## 5. Conclusion

By shifting the spectroscopic focus from average conductance channels to the intrinsic dynamics of current noise, this work establishes electronic noise spectroscopy as a highly sensitive, sub-thermal probe for molecular junctions. The experimental utilization of suspended-wire molecular junctions (SWMJs) functionalized with C10 and TPT demonstrates that peak-shaped vibrational features can be mapped with a level of resolution that bypasses the thermal line-width broadening typically limiting standard IETS. The exceptional detail captured by voltage-dependent current noise fluctuations validates the premise that electron transport at the nanoscale is heavily modulated by the fast scale time-dependence of molecular transmission as the system transitions between complex vibrational states.

The mathematical framework developed herein permits a deep parameterization of the junction's local landscape. This approach yields critical physical constants—including mode-specific anharmonicities, deep electronic potential dissociation energies, and localized outer-sphere reorganization energies —many of which have historically been inaccessible within mixed organic-metal structures.

Ultimately, the abundance of custom physical parameters extracted via this methodology delivers an intricate molecular signature. This dense multi-parameter profile provides a robust, unique fingerprint capable of discriminating between highly similar molecular species, paving a definitive path forward for the development of advanced quantum-scale chemical sensors.

## Acknowledgments

None

## Conflict of Interest Statement

The author Rani Arielly has no conflicts to disclose.

## Author Contributions

Rani Arielly: Conceptualization, Data curation, Formal analysis, Investigation, Methodology, Visualization, Writing – original draft, writing – review and editing;

## Data Availability Statement

The data that support the findings of this study are available from the corresponding author upon reasonable request.

**Figures:**

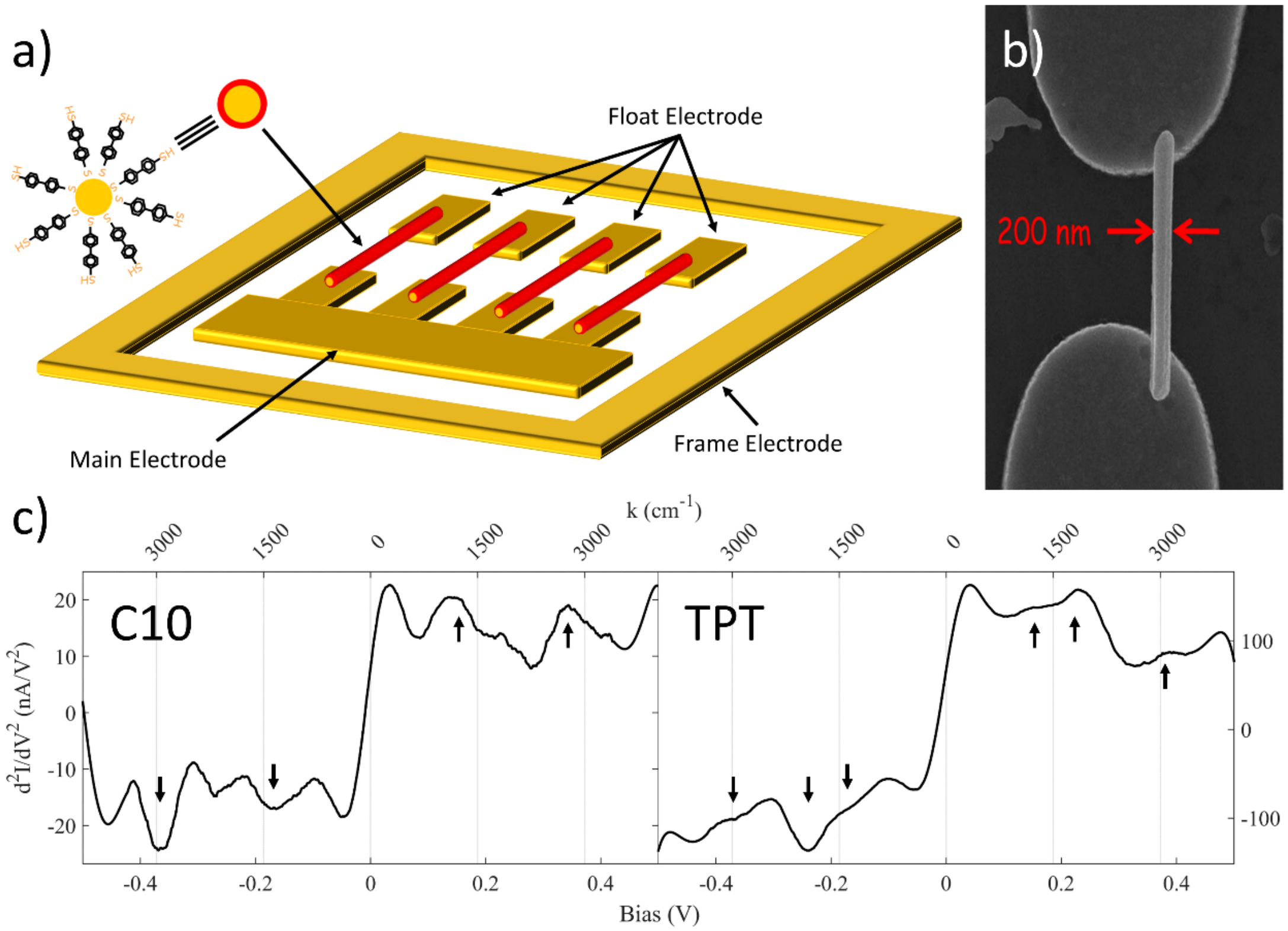

Figure 1. (a) Illustration of the SAM capped gold wire and the Au pads evaporated on the $Si/SiO_2$ substrate forming gaps which are bridged by the capped Au wires. (b) A close-up SEM image of the Au wire bridge area. The wires have a typical thickness of 200 nm and length of ~3 µm (Arielly et al., 2011). (c) IETS of junctions with C10 and TPT measured at $T_{LN}$ using an AC amplitude of 15 mV. Peaks of different characteristic vibrational modes are marked with arrows.

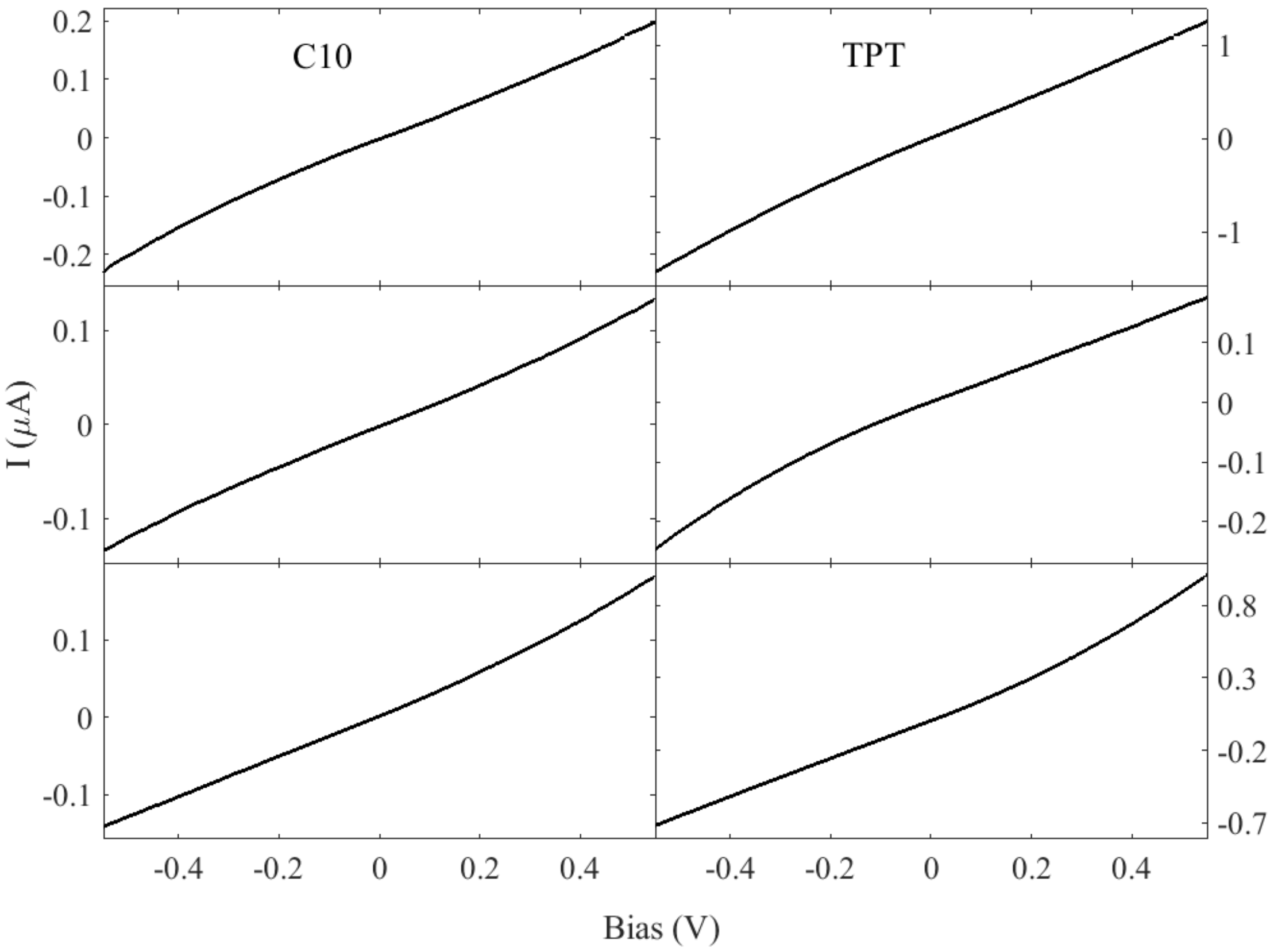

Figure 2. Typical $I(V)$ examples for both types of junctions. The left and right panels are for C10 and TPT junctions respectively.

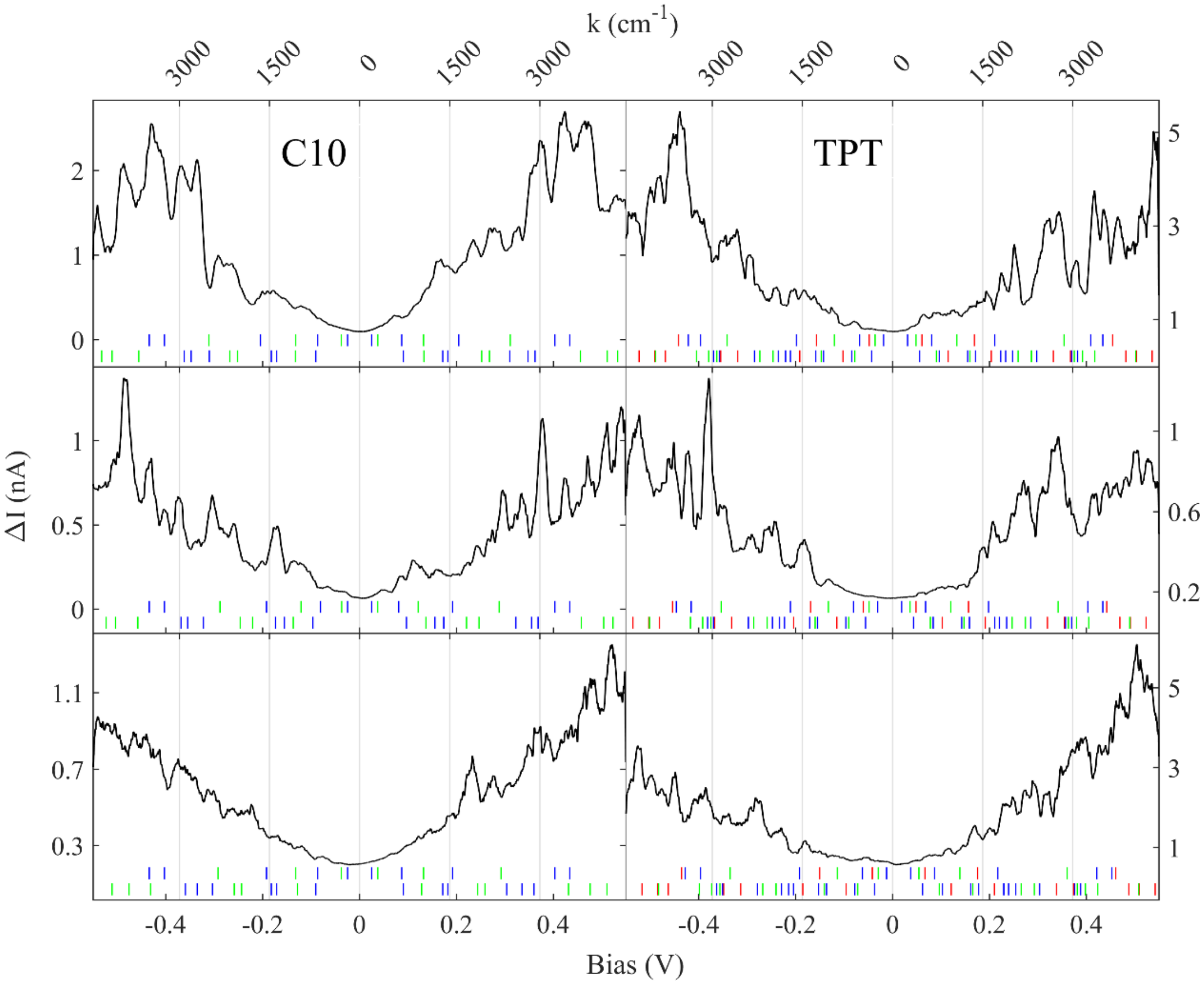

Figure 3. Typical $\Delta I(V)$ examples for both types of junctions. Peak onset is noted by the markers at the bottom of each graph which are separated vertically for different molecules (lower ones for C10/TPT and the upper ones for $H_2O$), and by color for base/overtones frequencies (blue is for the base frequency, green is for the first overtones, and red is for the second). The assignments are symmetrical in both bias polarities and give good coverage of the peak features. The left and right panels are for C10 and TPT junctions respectively.

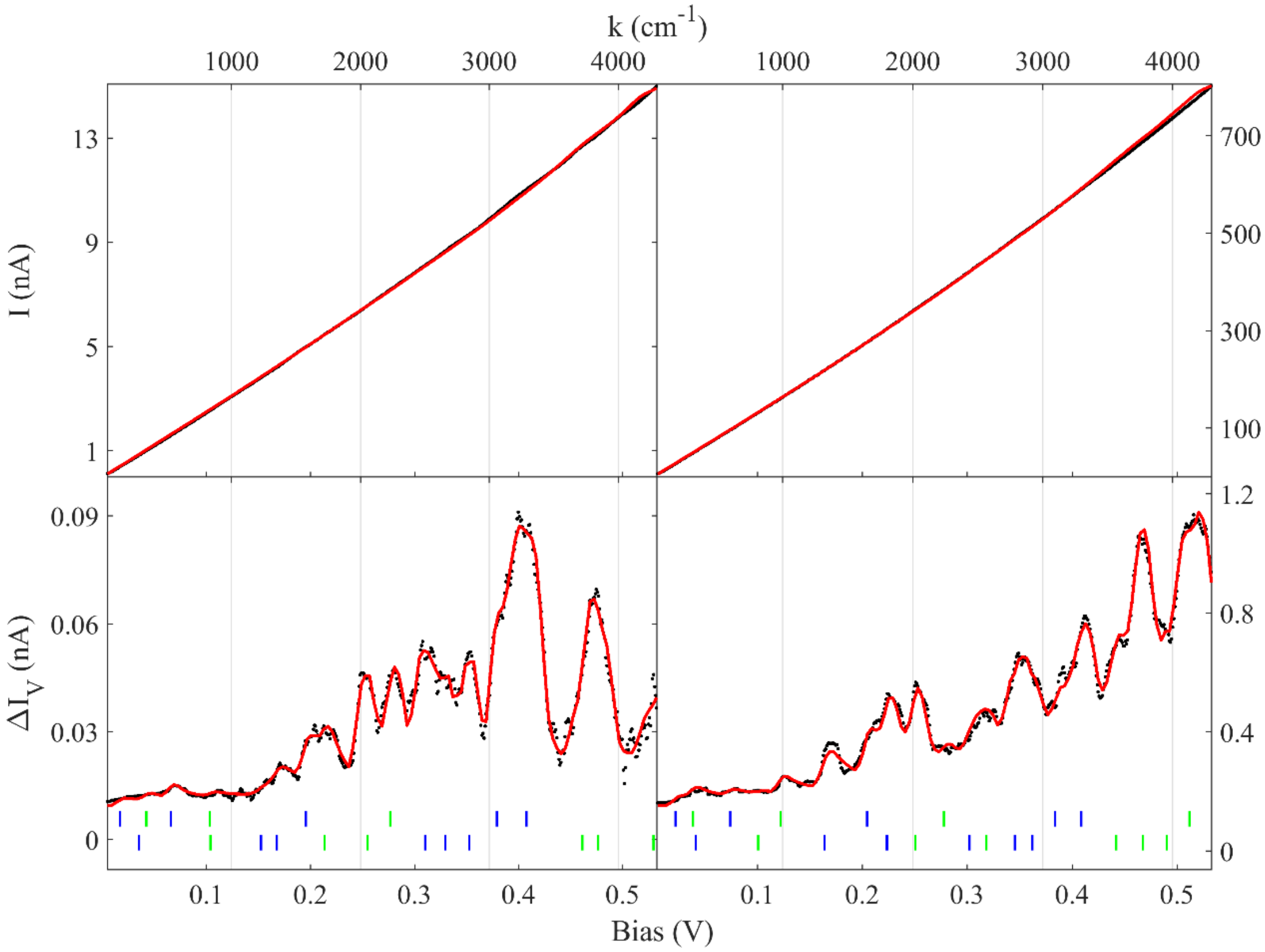

Figure 4. Fits of the model to $I(V)$ and $\Delta I_V(V)$ for 2 types of devices: C10 (left panes) and TPT (right panes). Black dots correspond to the harmonically averaged measured values, and red lines to the best fitted curves. Vibrational energies are noted by the markers at the bottom of each graph which are separated vertically for different molecules (lower ones for C10/TPT and the upper ones for $H_2O$), and by color for base/overtones frequencies (blue is for the base frequency and green is for the first overtones).

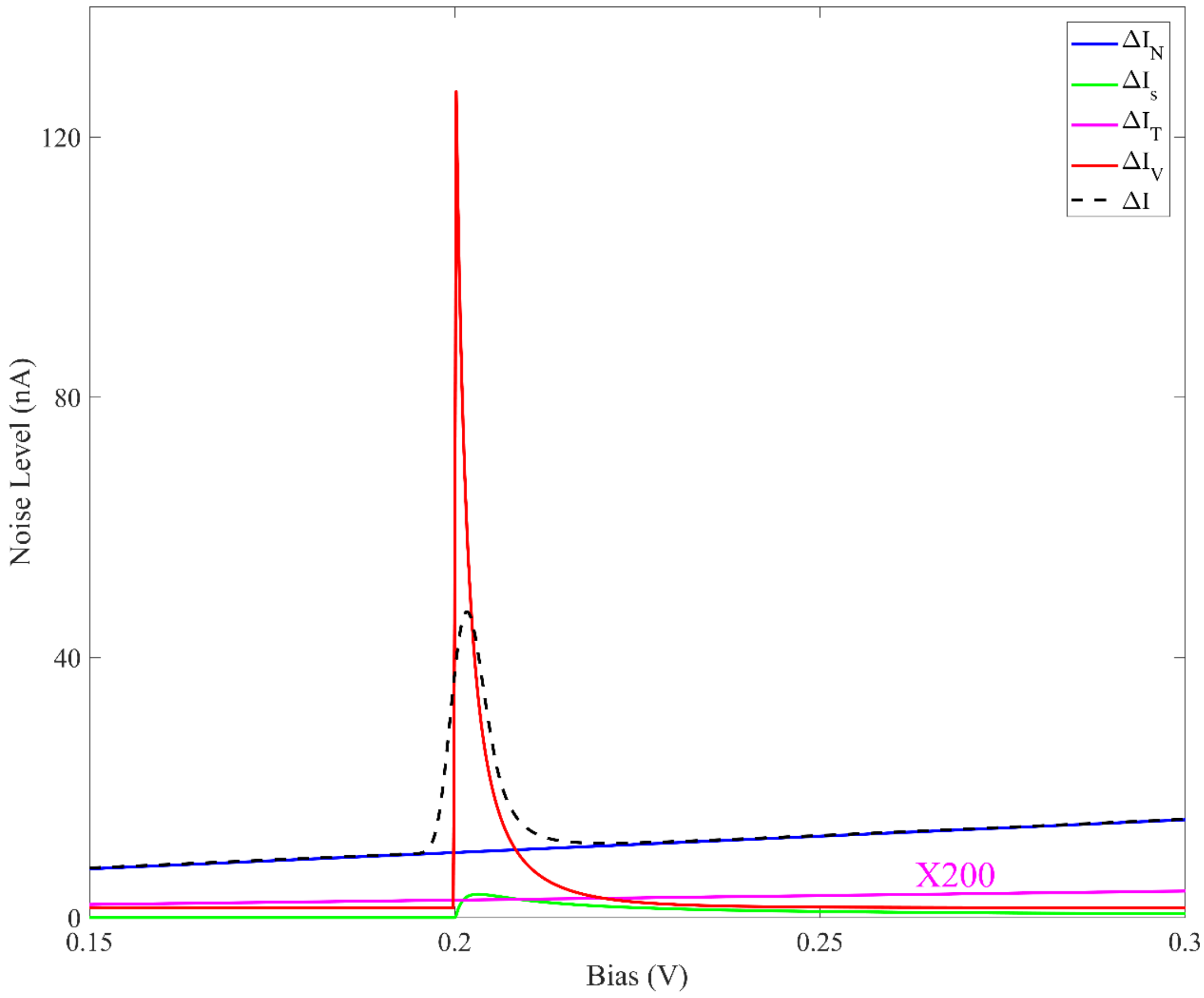

Figure 5. (a) Simulated noise components and total noise of a single molecular layer having only one vibration mode. The dominant terms are $\Delta I_N$ and $\Delta I_V$ when using parameters values as realistic as possible (see text). A peak feature is evident which originates from the $\Delta I_V$ component.

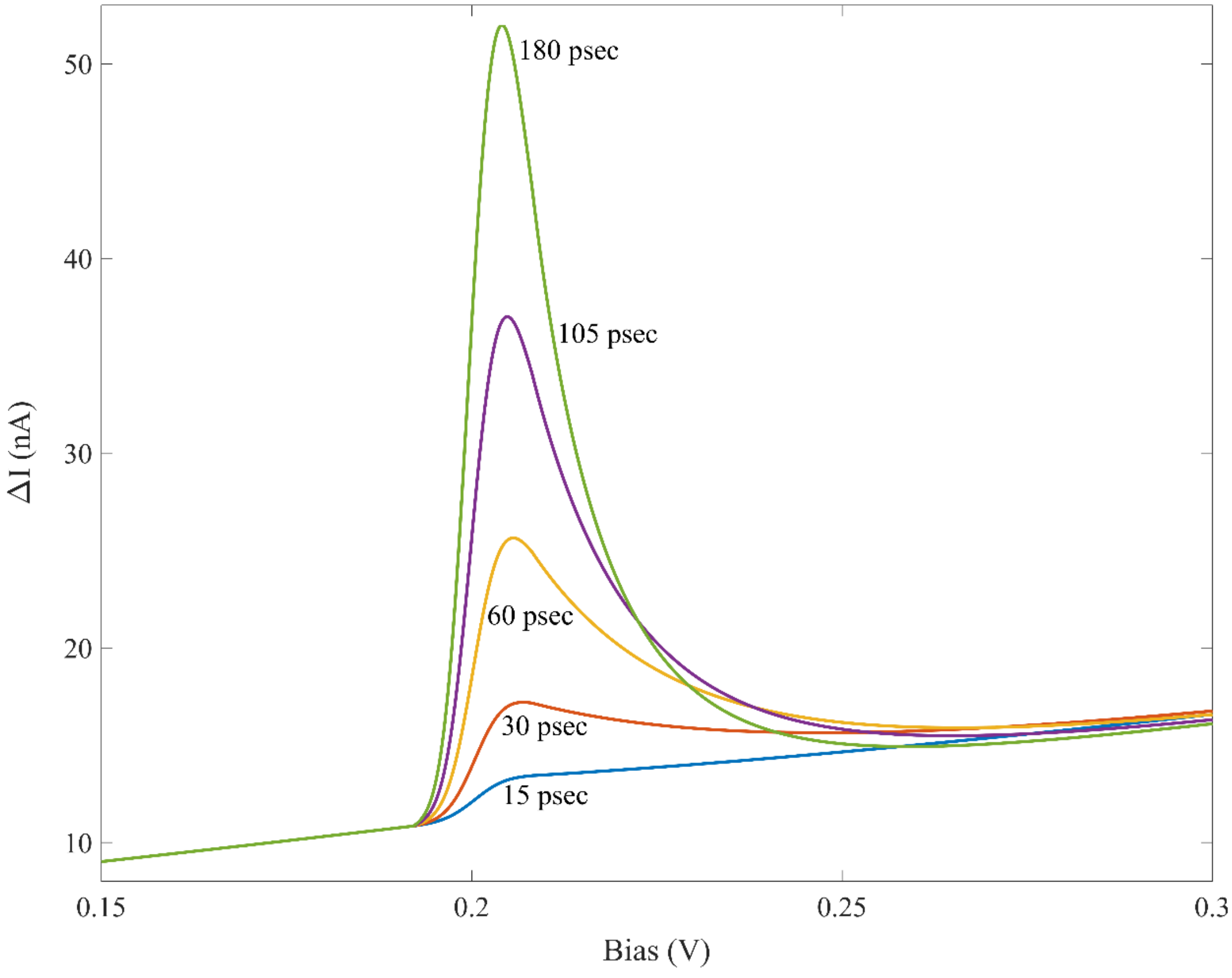

Figure 6. Effect of the $s$ parameter ($= \xi t^{decay}$) on the peak feature's line shape. For lower values, the peak feature becomes step like due to the inability to easily populate the excited mode once the bias threshold was reached.

Table I. The set of parameters that gave the best fits of the model to the $I(V)$ and $\Delta I_V(V)$ curves for 2 types of devices (C10 and TPT). Besides a general parameters list, individual lists exist for each phase, containing general parameters for the phase, and mode specific values. Notations include "st" for stretch, "sy" for symmetric, "as" for antisymmetric, "ar" for aromatic, δ for deformation vibration, γ for skeletal vibration, "oop" for out of plane, "comb" for combination band, "ot1" for first overtone, $v_2$ for H-O-H bending and $v_{Li}$ for $H_2O$ libration around the i axis.

| C10 | | | TPT | | |
|---|---|---|---|---|---|
| General parameters | | | | | |
| $N$ | 2.254(10) | | $N$ | 100.53(27) | |
| $\Delta V$ | 0.385 mV | | $\Delta V$ | 0.117 mV | |
| $V$ smoothing | 0.0159+0.0577(13) $V$ | | $V$ smoothing | 0.0159+0.04574(24) $V$ | |
| $\Delta I_0$ | 26.67(28) pA | | $\Delta I_0$ | 313.7(20) pA | |
| $\Delta N/N$ (%) | 0.598(4) | | $\Delta N/N$ (%) | 0.1080(6) | |
| Phase 1 | | | Phase 1 | | |
| $E_g$ | 1.7602(33) eV | | $E_g$ | 1.5917(14) eV | |
| $\gamma$ | 0.23335(43) eV | | $\gamma$ | 0.2303(2) eV | |
| $A$ | 0.0385(25) | | $A$ | 0.0262(12) | |
| Mode | $\nu$ (cm$^{-1}$) | $s$ (psec) | Mode | $\nu$ (cm$^{-1}$) | $s$ (psec) |
| $CH_2$ γ | 290(570) | 34(16) | ar C-H δ oop | 330(140) | 54(4) |
| $CH_2$ γ ot1 | 840(400) | 13.5(45) | ar C-H δ oop ot1 | 810(150) | 21.3(36) |
| $CH_3$ δ sy | 1230(100) | 51(13) | ar C-C | 1320(18) | 111(7) |
| $CH_3$ δ as | 1350(70) | 80(16) | comb | 1803(7) | 242(15) |
| $CH_3$ δ sy ot1 | 1726(20) | 242(42) | ar C-C ot1 | 2023(5) | 418(30) |
| $CH_3$ δ as ot1 | 2058(7) | 6000(2400) | S-H st | 2437(11) | 184(15) |
| S-H st | 2504(8) | 4100(1800) | comb ot1 | 2566(11) | 273(24) |
| -C-H st sy | 2660(12) | 1760(650) | ar C-H st as | 2787(7) | 1170(170) |
| -C-H st as | 2845(20) | 2500(900) | ar C-H st sy | 2921(6) | 1000(160) |
| S-H st ot1 | 3721(18) | 330(70) | S-H st ot1 | 3565(5) | 240(14) |
| -C-H st sy ot1 | 3844(23) | 1220(370) | ar C-H st as ot1 | 3772(1) | 1380(110) |
| -C-H st as ot1 | 4270(300) | 310(70) | ar C-H st sy ot1 | 3956(4) | 320(24) |
| Phase 2 | | | Phase 2 | | |
| $E_g$ | 0.6994(12) eV | | $E_g$ | 0.6951(6) eV | |
| $\gamma$ | 0.2579(5) eV | | $\gamma$ | 0.25592(25) eV | |
| $A$ | 0.0119(16) | | $A$ | 0.0116(9) | |
| Mode | $\nu$ (cm$^{-1}$) | $s$ (psec) | Mode | $\nu$ (cm$^{-1}$) | $s$ (psec) |
| $H_2O$ $v_{Lz}$ | 137(56) | 460(130) | $H_2O$ $v_{Lz}$ | 175(25) | 297(45) |
| $H_2O$ $v_{Lz}$ ot1 | 341(60) | 284(53) | $H_2O$ $v_{Lz}$ ot1 | 311(28) | 228(29) |
| $H_2O$ $v_{Ly}$ | 533(37) | 354(74) | $H_2O$ $v_{Ly}$ | 597(43) | 139(14) |
| $H_2O$ $v_{Ly}$ ot1 | 830(190) | 174(25) | $H_2O$ $v_{Ly}$ ot1 | 984(18) | 176(17) |
| $H_2O$ $v_2$ | 1580(23) | 350(56) | $H_2O$ $v_2$ | 1651(10) | 247(19) |
| $H_2O$ $v_2$ ot1 | 2233(10) | 574(78) | $H_2O$ $v_2$ ot1 | 2240(15) | 107(7) |
| O-H st sy | 3060(4) | 467(37) | O-H st sy | 3096(11) | 116(7) |
| O-H st as | 3288.4(21) | 1430(170) | O-H st as | 3299(5) | 211(12) |
| | | | O-H st sy ot1 | 4128(3) | 225(11) |

Table II. The base wavenumber ($\tilde{\nu}_e$), first order anharmonicity ($x_e$) and depth of the electronic potential ($D_e$) for several modes that exhibited overtones in our measurements. Mode notations are as in Table I.

| C10 | | | | TPT | | | |
|---|---|---|---|---|---|---|---|
| | $\tilde{\nu}_e$ (cm$^{-1}$) | $x_e$ | $D_e$ (cm$^{-1}$) | | $\tilde{\nu}_e$ (cm$^{-1}$) | $x_e$ | $D_e$ (cm$^{-1}$) |
| $CH_3$ δ sy | 1970(310) | 0.19(6) | 6200(2000) | ar C-C | 1937(53) | 0.159(10) | 7050(430) |
| $CH_3$ δ as | 2010(210) | 0.16(4) | 7200(1700) | comb | 2843(24) | 0.1829(36) | 9190(170) |
| S-H st | 3792(31) | 0.1698(36) | 13060(260) | S-H st | 3745(33) | 0.1747(34) | 12590(240) |
| -C-H st sy | 4136(42) | 0.1784(44) | 13660(320) | ar C-H st as | 4589(20) | 0.1964(17) | 13980(110) |
| -C-H st as | 4260(310) | 0.166(38) | 15000(3100) | ar C-H st sy | 4807(18) | 0.1961(15) | 14660(110) |
| $H_2O$ $v_{Ly}$ | 770(220) | 0.15(14) | 2900(2500) | $H_2O$ $v_{Ly}$ | 810(130) | 0.130(58) | 3500(1500) |
| $H_2O$ $v_2$ | 2507(71) | 0.185(11) | 8030(460) | $H_2O$ $v_2$ | 2712(32) | 0.1957(50) | 8290(200) |
| | | | | O-H st sy | 5159(32) | 0.2000(24) | 15480(180) |

Table III. Calculated reorganization energies for the different modes in our devices. Mode notations are as in Table I.

| | C10 | | TPT | |
|---|---|---|---|---|
| | Mode | $E_r$ (eV) | Mode | $E_r$ (eV) |
| Phase 1 | $CH_2$ γ | 0.329(65) | ar C-H δ oop | 0.349(14) |
| | $CH_2$ γ ot1 | 0.334(19) | ar C-H δ oop ot1 | 0.347(8) |
| | $CH_3$ δ sy | 0.389(9) | ar C-C | 0.416(2) |
| | $CH_3$ δ as | 0.406(6) | comb | 0.451(2) |
| | $CH_3$ δ sy ot1 | 0.449(6) | ar C-C ot1 | 0.472(2) |
| | $CH_3$ δ as ot1 | 0.558(13) | S-H st | 0.452(3) |
| | S-H st | 0.551(14) | comb ot1 | 0.466(3) |
| | -C-H st sy | 0.527(12) | ar C-H st as | 0.515(5) |
| | -C-H st as | 0.541(11) | ar C-H st sy | 0.512(5) |
| | S-H st ot1 | 0.484(7) | S-H st ot1 | 0.472(2) |
| | -C-H st sy ot1 | 0.527(10) | ar C-H st as ot1 | 0.530(3) |
| | -C-H st as ot1 | 0.486(7) | ar C-H st sy ot1 | 0.474(2) |
| Phase 2 | $H_2O$ $v_{Lz}$ | 0.389(16) | $H_2O$ $v_{Lz}$ | 0.383(7) |
| | $H_2O$ $v_{Lz}$ ot1 | 0.403(8) | $H_2O$ $v_{Lz}$ ot1 | 0.393(5) |
| | $H_2O$ $v_{Ly}$ | 0.424(7) | $H_2O$ $v_{Ly}$ | 0.398(4) |
| | $H_2O$ $v_{Ly}$ ot1 | 0.415(9) | $H_2O$ $v_{Ly}$ ot1 | 0.421(3) |
| | $H_2O$ $v_2$ | 0.458(5) | $H_2O$ $v_2$ | 0.449(2) |
| | $H_2O$ $v_2$ ot1 | 0.485(4) | $H_2O$ $v_2$ ot1 | 0.432(2) |
| | O-H st sy | 0.489(3) | O-H st sy | 0.444(2) |
| | O-H st as | 0.527(4) | O-H st as | 0.466(2) |
| | | | O-H st sy ot1 | 0.486(2) |